\pdfoutput=1

\documentclass[fleqn,12pt]{article}
\usepackage{natbib}
\usepackage{float,multirow}
\usepackage{amsthm,amsfonts,amsopn,amsmath,amssymb,verbatim}
\RequirePackage[pagebackref=true]{hyperref}
\usepackage{pgf,pgfplots,tikz}
\usetikzlibrary{arrows,calc,fit,matrix,positioning,shapes.multipart,shapes.symbols}
\usepackage{graphicx}
\usepackage{bbm}

\usepackage{natbib}
\usepackage{color}  
\usepackage{bm}
\theoremstyle{definition}
\newtheorem{theorem}{Theorem}
\newtheorem{proposition}{Proposition}
\newtheorem{example}{Example}

\def\dataSet{B}
\def\data{b}
\def\paramSet{A}
\def\param{a}
\def\mytimes{}

\hypersetup{colorlinks=true,linkcolor=blue,urlcolor=blue,citecolor=red,
    pdftitle=Comment on Estimating Dynamic Discrete Choice Models with Hyperbolic Discounting by Hanming Fang and Yang Wang,
    pdfauthor=Jaap Abbring and Oeystein Daljord,
    pdfsubject=JEL Codes C25 C61,
    pdfkeywords=discount factor dynamic discrete choice generic identification transversality theorem 
    pdfdisplaydoctitle=true}

\usepackage[nohead]{geometry} 
\usepackage[onehalfspacing]{setspace}

\title{A Comment on ``Estimating Dynamic Discrete Choice Models with Hyperbolic Discounting'' by Hanming Fang and Yang Wang%
\thanks{This comment incorporates material from Appendix B of the August 2018 version of \citet{ddc19:abbringdaljord} (\href{https://arxiv.org/abs/1808.10651v1}{arXiv:1808.10651v1} [econ.EM]), which we have deleted from the current draft of that paper. Thanks to Hanming Fang, Christian Hansen, and Eduardo Souza-Rodrigues for helpful comments and discussion. }%
}

\author{Jaap H. Abbring\thanks{CentER, Department of Econometrics \& OR, Tilburg University, P.O. Box 90153, 5000 LE Tilburg, The Netherlands; and CEPR. E-mail: \href{mailto:jaap@abbring.org}{jaap@abbring.org}. Web: \href{http://jaap.abbring.org}{jaap.abbring.org}.}\and
\O ystein Daljord\thanks{Booth School of Business, University of Chicago, 5807 South Woodlawn Avenue, Chicago, IL 60637, USA. E-mail:
\href{mailto:Oeystein.Daljord@chicagobooth.edu}{Oeystein.Daljord@chicagobooth.edu}. Web: \href{http://faculty.chicagobooth.edu/oystein.daljord}{faculty.chicagobooth.edu/oystein.daljord}. 
}}

\date{May 2019}

\begin{document}
\maketitle

\begin{abstract}
The recent literature often cites \citet{ier15:fangwang} for analyzing the identification of time preferences in dynamic discrete choice under exclusion restrictions \citep[e.g.][]{yaoetal12,Lee2013,ChingErdemKeane,res14:noretstang,dube2014joint,ms15:gordonsun,qme2016:bajarietal,res17:chan,qe18:gayleetal}. \citeauthor{ier15:fangwang}'s Proposition 2 claims generic identification of a dynamic discrete choice model with hyperbolic discounting. This claim uses a definition of ``generic'' that does not preclude the possibility that a generically identified model is nowhere identified. To illustrate this point, we provide two simple examples of models that are generically identified in \citeauthor{ier15:fangwang}'s sense, but that are, respectively, everywhere and nowhere identified. We conclude that Proposition 2 is void: It has no implications for identification of the dynamic discrete choice model. We show that its proof is incorrect and incomplete and suggest alternative approaches to identification. 
\end{abstract}

\pagebreak

\section{Introduction}

\citet{ier15:fangwang} studied the identification an infinite-horizon, stationary dynamic discrete choice model with  partially naive hyperbolic time preferences.
In each period, the agent chooses an action $i$ from ${\cal I}\equiv\{0,1,\ldots,I\}$, $I\in\mathbb{N}$, after she observes that period's Markov state $(x,\mathbf{\varepsilon})$, where $x$ takes values in a finite set ${\cal X}$ and $\mathbf{\varepsilon}=(\varepsilon_0,\varepsilon_1,\ldots,\varepsilon_I)\in\mathbb{R}^{I+1}$.\footnote{This paper's footnotes document various minor errors and inconsistencies in \citeauthor{ier15:fangwang} that are not central to our comments, but that we have corrected in the main text to ensure clarity and consistency. Here, for example, we have included $\varepsilon_0$ in $\varepsilon$. \citeauthor{ier15:fangwang} (p. 568) specified $\varepsilon=(\varepsilon_1,\ldots,\varepsilon_I)\in\mathbb{R}^{I}$ and only assumed $u^*_i(x,\mathbf{\varepsilon})=u_i(x)+\epsilon_i$ for $i\in{\cal I}/\{0\}$. However, \citeauthor{ier15:fangwang} subsequently used $u_0^*(x,\mathbf{\varepsilon})=u_0(x)+\varepsilon_0$, with $\varepsilon_0,\ldots,\varepsilon_I$ independent with type-1 extreme value distributions, to get logit choice probabilities $P_i(x)$.} This returns instantaneous utility $u_i^*(x,\mathbf{\varepsilon})=u_i(x)+\epsilon_i$. It also controls the evolution of $x$: Given choice $i$ in state $(x,\epsilon)$, it takes the value $x'\in{\cal X}$ in the next period with probability $\pi(x'|x,i)$. In contrast, the components of $\varepsilon'$ are mutually independent with type-1 extreme value distributions, independently from  $x'$, $(x,\varepsilon)$, and choice $i$. The agent has rational expectations; in particular, she believes $x$ to evolve according to the controlled Markov transition distribution $\pi$. She discounts future utility with a standard factor $\delta$ and present bias factor $\beta$, and perceives future selves to have present bias factor $\tilde\beta$. 
With a normalization $u_0(x)=0$ for all $x\in{\cal X}$, the model's unknown primitives are an $I\mytimes X$-vector $\mathbf{u}$ with the values of $u_i(x)$ for $i\in{\cal I}/\{0\}$ and $x\in{\cal X}$, the discount function parameters $(\beta,\tilde\beta,\delta)$, and a matrix $\mathbf{\Pi}$ with the state transition probabilities $\pi(x'|x,i)$ for $i\in{\cal I}$, $x\in{\cal X}$, and $x'\in{\cal X}$. Here, $X=|{\cal X}|$ is the number of elements of ${\cal X}$.

The econometrician's data are the state transition probabilities $\mathbf{\Pi}$ and a matrix $\tilde{\mathbf{P}}$ that collects the conditional probabilities $P_i(x)$ that the agents chooses $i$ in state $x$, $i\in{\cal I}$ and $x\in{\cal X}$. Because probabilities sum to one, the $(I+1)\mytimes X+(I+1)\mytimes X^2$ choice and transition probabilities in $\left(\tilde{\mathbf{P}},\mathbf{\Pi}\right)$ can be represented by a  vector that stacks $I\mytimes X+(I+1)\mytimes X\mytimes (X-1)$ of them. We adopt this representation and take $\left(\tilde{\mathbf{P}},\mathbf{\Pi}\right)\in[0,1]^{I\mytimes X+(I+1)\mytimes X\mytimes (X-1)}\subset\mathbb{R}^{I X+(I+1)X(X-1)}$.\footnote{\label{fn:probRep}\citeauthor{ier15:fangwang}'s online Appendix C instead specifies
\[
\left(\tilde{\mathbf{P}},\mathbf{\Pi}\right)\in\Delta^{(I+1)X}\times\left(\overbrace{\Delta^X\times\cdots\times\Delta^X}^{X\text{ copies}}\right)^{I+1}\subset\mathbb{R}^{I X+(I+1)X(X-1)}, 
\]
without defining $\Delta$. We guess that, for $J\in\mathbb{N}$, $\Delta^J\equiv\{(p_1,\ldots,p_J)\in\mathbb{R}^J:p_1\geq 0,\ldots,p_J\geq 0;\sum_{j=1}^Jp_j=1\}$ denotes a probability simplex, but then $\Delta^{(I+1)X}\times\left(\Delta^X\times\cdots\times\Delta^X\right)^{I+1}$ lies in a $I X+(I+1)X(X-1)$-dimensional linear subspace of  $\mathbb{R}^{(I+1) X+(I+1)X^2}$ rather than in $\mathbb{R}^{I X+(I+1)X(X-1)}$. However, all that matters for the reading of \citeauthor{ier15:fangwang}'s Proposition 2 and our comments is that \citeauthor{ier15:fangwang} use Lebesgue measure on $\mathbb{R}^{I X+(I+1)X(X-1)}$ to decide between generic and exceptional sets of data; see Footnote \ref{fn:generic}.}
The transition probabilities $\mathbf{\Pi}$ directly identify the agent's (rational) beliefs. The conditional choice probabilities $\tilde{\mathbf{P}}$ are linked to the model's primitives by the assumption that the agent's actions follow a  {\em stationary perception-perfection perfect strategy profile} of the decision problem with beliefs $\mathbf{\Pi}$ and some utilities $\mathbf{u}^*$ and discount factors $(\beta^*,\tilde\beta^*,\delta^*)$. 
The extreme-value assumption ensures that the conditional choice probabilities $P_i(x)$ have the logit form. As in the special case with geometric discounting ($\beta=\tilde\beta=1$), an application of \citet{res93:hotzmiller}'s choice probability inversion gives 
$I\mytimes X$ equations that relate the $I\mytimes X+3$ parameters $(\mathbf{u},\beta,\tilde\beta,\delta)$ to the data $(\tilde{\mathbf{P}},\mathbf{\Pi})$, one for each log choice probability contrast $\ln P_i(x)-\ln P_0(x)$, $i\in{\cal I}/\{0\}$ and $x\in{\cal X}$.\footnote{\citeauthor{ier15:fangwang} provided the analysis leading to these equations, but not the final equations themselves.} 

For its main identification result (Proposition 2), \citeauthor{ier15:fangwang} (p. 579) assumed that the observed state can be partitioned as $x=(x_r,x_e)$; where $x_r$ takes values in ${\cal X}_r$, $x_e$ takes values in ${\cal X}_e$, ${\cal X}={\cal X}_r\times{\cal X}_e$, and $|{\cal X}_e|\geq 2$; and that its Assumption 5 holds for all $(x_r,x_e)\in{\cal X}$ and $(x_r,x_e')\in{\cal X}$.\footnote{\citeauthor{ier15:fangwang} used the same notation for random variables and their realizations and, in Assumption 5,  incorrectly referred to the state's values $x_1$ and $x_2$ as ``state variables.''} The first part\footnote{The second part of Assumption 5 requires that transition probabilities $\pi(\cdot | x,i)$ for some choice $i$ differ between the same $(x_r,x_e)$ and $(x_r,x_e')$. This condition cannot possibly be necessary for \citeauthor{ier15:fangwang}'s Proposition 2 to be true, as it holds generically according to its definition of ``generic'' (see Section \ref{s:void}).} of its Assumption 5 then requires that
\begin{equation}
\label{eq:exclusion}
u_i(x_r,x_e)=u_i(x_r,x_e')\text{ for all }i\in{\cal I}/\{0\},~x_r\in{\cal X}_r,\text{ and }\left(x_e,x_e'\right)\in{\cal X}_e\times {\cal X}_e,
\end{equation}
which are $I\mytimes \left(|{\cal X}_e|-1\right)\mytimes |{\cal X}_r|$ different and nontrivial exclusion restrictions, one for each $i\in{\cal I}/\{0\}$, each $x_r\in{\cal X}_r$, and each of the $|{\cal X}_e|-1$ distinct pairs of subsequent $x_e$ and $x_e'$ in the (arbitrarily) ordered set ${\cal X}_e$.\footnote{\label{fn:eqCount}\citeauthor{ier15:fangwang}'s online Appendix C instead states that ``the data must also satisfy the additional $I\times|{\cal X}_e|\times|{\cal X}_r|$ equations requiring that $u_i(x_r,x_e)=u_i(x_r)$ for each $i\in{\cal I}/\{0\}$, each $x_e\in{\cal X}_e$ and each $x_r\in{\cal X}_r$.''  Its subsequent analysis fails to appreciate that these $I\times|{\cal X}_e|\times|{\cal X}_r|$ equations come with $I\times|{\cal X}_r|$ additional parameters  $u_i(x_r)$; $i\in{\cal I}/\{0\}$, $x_r\in{\cal X}_r$ (it concludes that the exclusion restrictions yield a system of ``$I\times X+I\times|{\cal X}_e|\times|{\cal X}_r|$ equations in $I\times X+3$ unknowns $(\mathbf{u},\beta,\tilde\beta,\delta)$''). Clearly, on balance, these $I\mytimes|{\cal X}_e|\mytimes|{\cal X}_r|$ equations only introduce $I\mytimes|{\cal X}_e|\mytimes|{\cal X}_r|-I\mytimes|{\cal X}_r|=I\mytimes \left(|{\cal X}_e|-1\right)\mytimes |{\cal X}_r|$ additional restrictions, as in our representation. Of course, these restrictions are simply the equalities in \eqref{eq:exclusion} that can be derived by differencing  \citeauthor{ier15:fangwang}'s  equations $u_i(x_r,x_e)=u_i(x_r)$ and $u_i(x_r,x_e')=u_i(x_r)$.}

Taken together, the $I\mytimes X$ constraints resulting from \citeauthor{res93:hotzmiller}'s choice probability inversion and those in \eqref{eq:exclusion} implied by the exclusion restrictions form a system of $I\mytimes X+I\mytimes\left(|{\cal X}_e|-1\right)\mytimes |{\cal X}_r|$ nonlinear equations in the $I\mytimes X+3$ parameters $(\mathbf{u},\beta,\tilde\beta,\delta)$ and the data $(\tilde{\mathbf{P}},\mathbf{\Pi})$.  \citeauthor{ier15:fangwang} denoted this system with
\begin{equation}
\label{eq:G}
\tilde{\cal G}\left(\mathbf{u},\beta,\tilde\beta,\delta;\left(\tilde{\mathbf{P}},\mathbf{\Pi}\right)\right)=0.
\end{equation}
The system of equations \eqref{eq:G} contains all the information linking the unknown parameters $(\mathbf{u},\beta,\tilde\beta,\delta)$ to the data $(\tilde{\mathbf{P}},\mathbf{\Pi})$ under the assumed exclusion restrictions in \eqref{eq:exclusion}. Therefore, \citeauthor{ier15:fangwang} studied their model's identification by analyzing whether \eqref{eq:G} uniquely determines $(\mathbf{u},\beta,\tilde\beta,\delta)$ for given data. It claimed the following result:\footnote{We quote \citeauthor{ier15:fangwang}'s Proposition 2 verbatim, except that we have replaced its condition $I\mytimes |{\cal X}_e|\mytimes |{\cal X}_r|\geq 4$ with the stronger condition $I\mytimes \left(|{\cal X}_e|-1\right)\mytimes |{\cal X}_r|\geq 4$. Proposition 2's proof in \citeauthor{ier15:fangwang}'s online Appendix C relies on the fact that ``$I\times X + I\times|{\cal X}_e|\times |{\cal X}_r|$ ... is larger than the number of unknowns $I\times X + 3$ under our identifying assumption that $I\times |{\cal X}_e|\times |{\cal X}_r|\geq 4$.'' However, as we have explained in Footnote \ref{fn:eqCount}, the number of equations equals  $I\mytimes X+I\mytimes\left(|{\cal X}_e|-1\right)\mytimes |{\cal X}_r|$, not $I\mytimes X+I\mytimes|{\cal X}_e|\mytimes |{\cal X}_r|$, so that $I\mytimes \left(|{\cal X}_e|-1\right)\mytimes |{\cal X}_r|\geq 4$ is required to ensure that there are more equations than unknowns. Note that this correction neither changes the substance of \citeauthor{ier15:fangwang}'s proof, which simply relies on having more equations than unknowns, nor that of our comment.}
\begin{quote}
\vspace*{-7mm}
\setcounter{proposition}{1}
\begin{proposition}
Consider the space of data sets that can be generated by the assumed data generating process for some primitives $(\mathbf{u}^*,\beta^*,\tilde{\beta}^*,\delta^*)$. Suppose that there exist state variables that satisfy Assumption 5. Then, all the model parameters are generically identified if $I\mytimes \left(|{\cal X}_e|-1\right)\mytimes |{\cal X}_r|\geq 4$.
\end{proposition}
\end{quote}

\noindent \citeauthor{ier15:fangwang} does not formally define ``generic identification,'' but paraphrases Proposition 2 as giving identification ``for almost all data sets generated by the assumed hyperbolic discounting model'' (p. 579). \citeauthor{ier15:fangwang}'s proof of Proposition 2, in its online Appendix C, further defines ``almost all'' and therewith ``generic.'' 

The proof of Proposition 2 applies the transversality theorem to \citeauthor{ier15:fangwang}'s model to demonstrate that there are generically no parameters that are consistent with any given data. Next, it notes that since the model generated the data by assumption, there must exist some parameters consistent with the data. It concludes that these parameters are therefore generically the only parameters that are consistent with such data.

Section \ref{s:void} uncovers \citeauthor{ier15:fangwang}'s definition of ``generic'' from this proof. Then, it demonstrates that Proposition 2 is {\em void}: The model may be generically identified, in the sense of \citeauthor{ier15:fangwang}, independently of whether any data sets that can be generated by the model correspond to a unique parameter vector. It then shows that the proof is {\em incorrect}. Finally, it notes that the proof is {\em incomplete}, as it fails to verify the rank condition for the transversality theorem that it invokes. It is shown that independently of whether this rank condition holds, the proof has no implications for the model's identification. Section \ref{s:concl} concludes with a brief discussion of alternative approaches to identification in dynamic discrete choice models.

\section{A void generic identification result}
\label{s:void}

The proof of Proposition 2 in \citeauthor{ier15:fangwang}'s online Appendix C first presents the following transversality theorem \citep[][Proposition 8.3.1]{mascolell85}:\footnote{To avoid confusion with \citeauthor{ier15:fangwang}'s use of $x$ for states, we slightly deviate from \citeauthor{mascolell85}'s and \citeauthor{ier15:fangwang}'s notation and use $\param$ instead of $x$ and $\paramSet$ instead of $N$ here.}
\begin{theorem}[] 
Let $F : \paramSet\times \dataSet \rightarrow \mathbb{R}^m$, $\paramSet\subset\mathbb{R}^n$, $\dataSet\subset\mathbb{R}^s$ be $C^r$ with $r>\max\{n-m,0\}$. Suppose that $0$ is a regular value of $F$; that is, $F(\param,\data) = 0$ implies $\mathrm{rank}~ \partial F(\param,\data) = m$. Then, except for a set of $\data\in \dataSet$ of Lebesgue measure zero, $F_\data : \paramSet \rightarrow\mathbb{R}^m$ has $0$ as a regular value.
\end{theorem}
\noindent Here, $\partial F$ is the Jacobian of $F$ with respect to $(\param,\data)$; $F_\data$ is such that $F_\data(\param)=F(\param,\data)$ for all $\param,\data$; and $\partial F_\data$ is the Jacobian with respect to $\param$ only. 

To prove Proposition 2, it applies this transversality theorem to the system of equations \eqref{eq:G}, with the following mapping of notation:\footnote{This mapping corrects two minor problems with \citeauthor{ier15:fangwang}'s mapping at the top of page 3 of its online Appendix. See Footnotes \ref{fn:probRep} and \ref{fn:eqCount}.}
\begin{table}[h]
\centering
\begin{tabular}{ll}
\hline\hline
Transversality Theorem & \citeauthor{ier15:fangwang}\\
\hline
$F(\param,\data)\in\mathbb{R}^m$ & $\tilde{\cal G}\left(\mathbf{u},\beta,\tilde\beta,\delta;\left(\tilde{\mathbf{P}},\mathbf{\Pi}\right)\right)\in\mathbb{R}^{I\mytimes X + I\mytimes \left(|{\cal X}_e|-1\right)\mytimes |{\cal X}_r|}$\\
$m$&Number of equations in $\tilde{\cal G}$: $I\mytimes X + I\mytimes \left(|{\cal X}_e|-1\right)\mytimes |{\cal X}_r|$\\
$\param\in\paramSet\subset\mathbb{R}^n$&Unknown parameters $\left(\mathbf{u},\beta,\tilde\beta,\delta\right)\in\mathbb{R}^{I\mytimes X}\times(0,1]^3\subset\mathbb{R}^{I\mytimes X+3}$\\
$n$&Number of unknown parameters: $I\mytimes X+3$\\
$\data\in\dataSet\subset\mathbb{R}^s$&Vector of probabilities in $[0,1]^{I\mytimes X+(I+1)\mytimes X\mytimes (X-1)}\subset\mathbb{R}^{I\mytimes X+(I+1)\mytimes X\mytimes (X-1)}$\\
&~~~ that represents the data $\left(\tilde{\mathbf{P}},\mathbf{\Pi}\right)$\\
$s$&$I \mytimes X+(I+1)\mytimes X\mytimes (X-1)$\\
$\param\mapsto F_\data(\param)$&$\left(\mathbf{u},\beta,\tilde\beta,\delta\right)\mapsto\tilde{\cal G}\left(\mathbf{u},\beta,\tilde\beta,\delta;\left(\tilde{\mathbf{P}},\mathbf{\Pi}\right)\right)$\\
\hline 
\end{tabular}
\end{table}

\noindent That is, \citeauthor{ier15:fangwang} studied the generic identification of the vector $\param\in \paramSet\subset\mathbb{R}^n$ of unknown parameters $\left(\mathbf{u},\beta,\tilde\beta,\delta\right)$ from the choice and transition probabilities $\data\in \dataSet\subset\mathbb{R}^s$ by applying the transversality theorem to the system $F(\param,\data)=0$ of $m$ smooth equality constraints.  This implicitly defines  ``for almost all data sets generated by the assumed hyperbolic discounting model'' (and therewith ``generically'' in Proposition 2) to mean for all data $\data\in\dataSet$ in the model's range  (the set of $\data\in\dataSet$ such that $F(\param,\data)=0$ has at least one solution $\param\in\paramSet$) outside a set of Lebesgue measure zero in  $\mathbb{R}^s$.\footnote{\label{fn:generic}Recall from Footnote \ref{fn:probRep} that it is not completely clear how \citeauthor{ier15:fangwang} represent the choice and transition probability data, but that it is clear that they think of the data as living in $\mathbb{R}^s=\mathbb{R}^{I X+(I+1)X(X-1)}$. The exact way the data are represented in $\mathbb{R}^s$ is irrelevant, because Lebesgue measure is invariant under affine transformations with determinant $1$ or $-1$. In particular, \citeauthor{ier15:fangwang}'s representation and ours both assign zero measure to the same sets of choice and transition probabilities.}   

A key problem with \citeauthor{ier15:fangwang}'s Proposition 2 is that its proof, given that the assumed rank condition holds (we return to this at the end of this section), establishes that the model's range has Lebesgue measure zero in $\mathbb{R}^s$. Because the transversality theorem, as applied in \citeauthor{ier15:fangwang}'s proof, only has implications for data outside a set of Lebesgue measure zero, it has no consequences for identification from data in the model's range.  

To be precise, suppose that the rank condition for the transversality theorem holds: $\mathrm{rank}~\partial  F(\param,\data)=m$ if $F(\param,\data)=0$. Then, the transversality theorem implies that,  for all $\data\in\dataSet$ outside a set of Lebesgue measure zero in $\mathbb{R}^s$, $\mathrm{rank}~\partial  F_\data(\param)=m$ if $F(\param,\data)=0$. Moreover, because $\param\in\mathbb{R}^n$, $\mathrm{rank}~\partial  F_\data(\param)\leq n<m$. Taken together, this implies that $F(\param,\data)=0$ has no solutions $\param\in\paramSet$, except for $\data\in\dataSet$ in a set of Lebesgue measure zero in $\mathbb{R}^s$. 

Consequently, given that the rank condition holds, Proposition 2 is void. It claims that $F(\param,\data)=0$ has a unique solution $\param\in\paramSet$ for all $\data\in\dataSet$ in the model's range. Since the model's range has zero measure, it is excepted from the claim. Proposition 2 therefore makes no claim about the number of solutions in the range of the model. We note that Proposition 2 is not false. Formally, Proposition 2 is vacuously true, because it is a statement about a property of the elements of an empty set.\footnote{It is vacuously true since any statement about a property of elements of an empty set is formally true.}

Moreover, its proof cannot easily be adapted to establish a more substantive identification result, for some or all data in the model's range, because its application of the transversality theorem has no implications for the number of parameters $\param\in\paramSet$ that solves $F(\param,\data)=0$ for data $\data\in\dataSet$ in the model's zero measure range.

We illustrate these two points with two simple examples. We first note that \citeauthor{ier15:fangwang}'s proof does not use the particular structure of the dynamic discrete choice model, but applies to any model that can be represented by a system of equations with more equations than unknowns under the regularity conditions stated above. Our examples therefore use highly stylized, linear models that allow easy and direct verification of the rank condition and the conclusions of the transversality theorem. Like \citeauthor{ier15:fangwang}'s model under the conditions of Proposition 2, both examples have models with more equations than unknowns ($m>n$).  Their ranges have Lebesgue measure zero in $\mathbb{R}^s$, so that generic identification vacuously holds. However, in the first example, $\param$ is uniquely determined from $F(\param,\data)=0$ for $\data\in B$ in the model's range; in the second example, it is not. 
%
\begin{example}[\bf Everywhere point identified]

Suppose that the data are $\data = (\data_1, \data_2)\in \dataSet=\mathbb{R}^2$, the parameter is $\param \in \paramSet=\mathbb{R}$, and the model is $F:\mathbb{R} \times \mathbb{R}^2 \rightarrow \mathbb{R}^2$, with
\[
0 = F(\param;\data) =
\left[	
\begin{array}{c}
\data_1 - \param \\
\data_2 - \param
\end{array}
\right],~~
\partial F(\param;\data) = 
\left[
\begin{array}{ccc}	
-1 &	1&	0 \\
-1 &	0 &	1
\end{array}
\right],\text{  and  }
\partial F_{\data}(\param) = 
\left[
\begin{array}{c}
-1 \\
-1
\end{array}
\right].
\]
Note that $n=1$, $s=2$, and $m=2$. In this example, $\mathrm{rank}~ \partial F(\param,\data) = 2$ always. The transversality theorem gives that $F_{\data}(\param)=0$ implies $\mathrm{rank}~ \partial F_{\data}(\param)=2$ for almost all $\data\in \mathbb{R}^2$. Now, $\mathrm{rank}~ \partial F_{\data}(\param)\leq n=1$, so $F_{\data}(\param) \neq 0$ for almost all $\data \in \mathbb{R}^2$. The model is linear, so we can do without the transversality theorem and directly observe that the model can only generate data $\data$ such that $\data_1=\data_2$, which is nongeneric in $B = \mathbb{R}^2$. Data $\data\in\dataSet$ that can be generated by this model uniquely determine $\param$.  In this example, the transversality theorem tells us that there are zero (not one) parameters that rationalize the data, for almost all data in $\mathbb{R}^2$, and that data that are in the model's range, excepted from the transversality theorem, {\em always} identify the unknown parameter.
\end{example}

\begin{example}[\bf Nowhere point identified]

Now suppose we have data $\data = (\data_1, \data_2, \data_3)\in\dataSet= \mathbb{R}^3$, a pair of parameters $\param = (\param_1,\param_2) \in\paramSet= \mathbb{R}^2$ , and a model is $F:\mathbb{R}^2 \times \mathbb{R}^3 \rightarrow \mathbb{R}^3$, with 
\[
0 = F(\param;\data) =
\left[	
\begin{array}{c}
\data_1 - (\param_1+ \param_2) \\
\data_2 - (\param_1+ \param_2) \\
\data_3 - (\param_1+ \param_2)
\end{array}
\right],
\]
\[ \partial F(\param;\data) = 
\left[
\begin{array}{ccccc}
-1&	-1	&1&0&	0 \\
-1&	-1 &	0&1&	0 \\
-1 & 	-1	&0	&0	&1
\end{array}
\right],\text{  and  }
 \partial F_{\data}(\param) = 
\left[	
\begin{array}{cc}
-1	&-1 \\
-1	&-1 \\
-1	&-1
\end{array}
\right].
\]
Note that $n=2$, $s=3$, and $m=3$. In this example, $\mathrm{rank}~ \partial F(\param,\data) = 3$ always. The transversality theorem gives that $F_{\data}(\param)=0$ implies that $\mathrm{rank}~ \partial F_\data(\param)=3$ for almost all $\data\in \mathbb{R}^3$. Since $\mathrm{rank}~ \partial F_{\data}(\param)\leq n=2$, $F_{\data} (\param)\neq 0$ for almost all $\data$. This again makes sense: The model, which requires $F(\param,\data) = 0$, can only generate data $\data$ such that $\data_1 = \data_2 = \data_3$, which is nongeneric in $\mathbb{R}^3$. Data from the range of the model only identify $\param_1+ \param_2$ and never $\param_1$ and $\param_2$ separately. So, this is another example where transversality tells us there are zero (not one) parameters that match the data for almost all data. However, unlike in the previous example, data in the model's range {\em never} identify the parameters.
\end{example}

Together, these examples show that the transversality theorem, as applied in \citeauthor{ier15:fangwang}'s proof, has no implications for identification. Given that the rank condition for its application of the transversality condition holds, \citeauthor{ier15:fangwang}'s proof is correct up to its last half sentence. The first half of the proof's last sentence correctly concludes that, for all data $\data\in\dataSet$ outside a set of Lebesgue measure zero, there exist no parameters $\param\in\paramSet$ that solve $F(\param,\data)=0$. However, the last half sentence qualifies this conclusion with ``except the true primitives $(\mathbf{u}^*,\beta^*,\tilde\beta^*,\delta^*)$... that generated the data.''  This qualification does not follow from the preceding mathematical arguments. In particular, we have shown that  \citeauthor{ier15:fangwang}'s application of the transversality theorem implies that zero, not one, parameters solve the model for all data outside a set of measure zero. The transversality theorem has no implications for the number of parameters that solve the model for data in an exceptional set, which includes the model's range. So, this last half sentence of  \citeauthor{ier15:fangwang}'s proof is incorrect.

Finally, \citeauthor{ier15:fangwang}'s proof is also incomplete, because it fails to verify the key rank condition for its application of the transversality theorem:  $\mathrm{rank}~\partial  F(\param,\data)=m$ if $F(\param,\data)=0$. Instead, \citeauthor{ier15:fangwang} noted that ``this can be verified in the same way that we verify [a similar condition in the proof of Proposition 1],'' but did not verify the latter condition either (p.578). The incompleteness of the proof is however immaterial for the conclusion that can be drawn from the proof. If the rank condition holds, we know that the model's range has Lebesgue measure zero and is excepted from the transversality theorem. If the rank condition is violated, the transversality theorem does not apply. Either way, the proof has no implications for identification.

\section{Discussion}
\label{s:concl}

The source of the problems with \citeauthor{ier15:fangwang}'s Proposition 2 is its focus on identification that is generic in the data space, rather than the parameter space. This is nonstandard and complicates the analysis in two ways. First, the specification of an appropriate measure directly on the data requires  knowledge of the model's empirical content, i.e. the range of data that can be generated by varying the model parameters on their domain. Our discussion of Proposition 2 highlights the problems of ignoring the model's empirical content. 

Second, it is unclear how the concept of generic identification in the data space corresponds to the concept of generic identification in the parameter space.  The two concepts are generally not interchangeable, as the following stylized example illustrates. Consider a model that maps a parameter $\theta\in\mathbb{R}$ to a choice probability $p=p(\theta)\in[0,1]$. Define "for almost all" $\theta$ (or $p$) to mean for all $\theta$ (or $p$) outside a set of Lebesgue measure $0$. If $p(\theta)=1/(1+\exp(\theta))$, then $\theta$ is identified for almost all $p$ and almost all $\theta$.  If instead $p(\theta)$ equals $0$ if $\theta\leq 0$, $\theta$ if $\theta\in(0,1)$, and $1$ if $\theta\geq 1$, then $\theta$ is identified for almost all $p$, but we not for almost all $\theta$. 

One could possibly derive an identification result for the case with more equations than free parameters that is generic in the parameter space instead, following e.g. \citeauthor{ecta83:sargan}, \citet{jem92:mcmanus}, and \citeauthor{jpe04:ekelandetal}. One would also have to choose between a measure-theoretic definition of ``genericity'', like \citeauthor{ier15:fangwang}'s, and a topological one. \citeauthor{jem92:mcmanus} and \citeauthor{jpe04:ekelandetal} provide discussion. Generic identification, however, is a weak concept of identification, and particularly so if the exceptional set cannot be characterized. The very small subsets where identification fails may happen to contain economically important models. One example is \citet{jpe04:ekelandetal} which shows that the generic identification of the hedonic model does not cover the linear-quadratic special case that is at the center of most applied work.

 \citeauthor{ddc19:abbringdaljord} offers an alternative approach that dispenses with the concept of generic identification. It instead exploits the specific structure of the dynamic discrete choice model to analyze identification of a special case of \citeauthor{ier15:fangwang}'s model, with exponential discounting ($\beta=\tilde\beta=1$). It shows that each exclusion restriction in \eqref{eq:exclusion} for some $x_r\in{\cal X}_r$ and distinct $x_e\in{\cal X}_e$ and $x_e'\in{\cal X}_e$ gives a single moment condition that relates the discount factor $\delta$ to the choice and transition probabilities.\footnote{The exclusion restrictions in \eqref{eq:exclusion} are special cases of the ones considered in \citeauthor{ddc19:abbringdaljord}.} These moment conditions contain all the information in the data about $\delta$, and therewith $\mathbf{u}$.\footnote{\citeauthor{ddc19:abbringdaljord}'s Section 4 noted that  a version of \citeauthor{ecta02:magnacthesmar}'s Proposition 2 holds: There exist unique (up to a standard utility normalization) values of the primitives (notably, $\mathbf{u}$) that rationalize the data for any given discount factor $\beta\in[0,1)$. The joint identification of a non-parametric utility function and the discount factor is therefore reduced to the conditions $\beta$ derived from the exclusion restriction.} The analysis shows that each single exclusion restriction in general gives set identification, where the identified set is finite. For important special cases, such as models with one-period finite dependence  \citep[e.g.][]{ecta87:rust,degrooteverboven18}, the exclusion restriction gives point identification. \citet{abbringdaljordiskhakov18} showed that it is similarly possible to concentrate the identification analysis of a model with sophisticated present biased preferences ($\tilde{\beta} = 1$) on a small number of moment conditions derived directly from equally many exclusion restrictions. However, this approach does not extend to \citeauthor{ier15:fangwang}'s partially naive case. The identification of partially naive time preferences seems to require an analysis of the full system of equations and remains an open question. 
  
\pdfbookmark[0]{References}{pdfbm:refs}
\bibliographystyle{chicago}
\bibliography{alljaap2}

\end{document}